\documentclass{article}
\usepackage{color}  
\usepackage{epsf}
\usepackage{amsmath,amssymb,graphicx}
\usepackage{bm}

\begin{document}

\renewcommand{\theequation}{\thesection.\arabic{equation}}

\title{Dynamo and the Adiabatic Invariant}

\author{Alexander M. Balk\thanks{balk@math.utah.edu}, \\Department of Mathematics, \\University of Utah, Salt Lake City, UT 84112}
\maketitle
\date
\begin{abstract}\noindent
The paper considers dynamo generated by a shallow fluid layer in a celestial body (planet or star).
This dynamo is based on the extra invariant for interacting magnetic Rossby waves. The magnetohydrodynamics (MHD) is linearized on the background of strong toroidal magnetic field. The extra invariant is used to show that the background field is maintained.
\end{abstract}

\section{Introduction}
\setcounter{equation}{0}
\label{Sect: Intro}

Magnetic field of a celestial body often crucially depends on the shallow fluid layer. In the case of Sun, this layer is tachocline \cite{SpieZahn}, although its role is currently questioned, see review \cite{Charbonneau20}. In the case of Earth, researchers often assume the existence of stably stratified layer at the top of the liquid iron outer core, the stratified ocean of the core in terminology of \cite{Brag98}. We consider magnetohydrodynamics (MHD) of such layer locally in the plane 3D layer tangent to the planet or star. 

We linearize this dynamics on the background of strong toroidal magnetic field $B_0$ and note the possibility to separate variables for an arbitrary stratification. This leads to the quadratic dispersion relation, characterized by two length scales:
\begin{eqnarray}\label{rl}
{\sf r}=c_g/f\qquad \& \qquad \ell=\sqrt{|B_0/\beta|}.
\end{eqnarray} 
The scale ${\sf r}$ is the Rossby radius of deformation; it characterizes the usual hydrodynamic Rossby waves without magnetism; $f$ is the angular speed of local rotation; $c_g=\sqrt{g h}$ is the gravity wave speed, where the height $h$ is determined by the eigen-value problem resulting from the separation of variables.
The scale $\ell$ characterizes the strength of the background magnetic field $B_0$; $\beta$ is the derivative of $f$ with respect to latitudinal variable.

We notice that in each of the two cases (Earth and Sun), these scales have the same order of magnitude. If $\ell={\sf r}$ exactly, the quadratic dispersion relation can be factorized, resulting in the two dispersion laws that are exactly
\begin{subequations}\label{magneticAlfven}
\begin{eqnarray}
\omega_A&=&-|B_0| p, \\
\omega_m&=&\frac{|B_0| p\, {\sf r}^2 k^2}{1+{\sf r}^2 k^2}
\end{eqnarray}
\end{subequations}
(Alfven and magnetic Rossby waves); ${\bf k}=(p,q)$ is the wave vector ($k^2=p^2+q^2)$.

We then see that the long magnetic Rossby waves, (\ref{magneticAlfven}b), mostly carry magnetic energy, while in short waves, the mechanical and magnetic energies have roughly the same magnitude. The following picture emerges. 
Some source (e.g. the arrival of light fluid bubbles from the inner core) excites relatively short waves (${\sf r}k\gtrsim1$). Interacting, they produce other waves. Most of the energy from the source is transferred towards large $k$, but a small fraction of that energy gets into long waves (${\sf r} k\ll1$) and accumulates there. Most of the accumulated energy is magnetic.

We use the adiabatic invariant \cite{BNZ,B1991} to show  that the energy mainly accumulates in zonal magnetic field, and so, the background field $B_0$ is maintained. This extra invariant was found for the usual hydrodynamic Rossby waves and plasma drift waves. But since the dispersion law (\ref{magneticAlfven}b) differs from the well known Rossby dispersion law only by a Doppler shift, the extra invariant holds for the magnetic Rossby waves as well. It has an unusual spectral density
\begin{eqnarray}
\eta=\arctan\frac{q+p\sqrt{3}}{{\sf r}k^2}-\arctan\frac{q-p\sqrt{3}}{{\sf r}k^2}\label{ETA};
\end{eqnarray}
the invariant equals integrated over ${\bf k}$ the product of this function with the wave action spectrum, 
Section \ref{Sect: Dynamo}.  The function (\ref{ETA}) is actually the only density that gives an independent extra invariant (in addition to the energy and momentum). 
We then find the proportion of magnetic energy contained in different components of magnetic field. Finally, we consider the  build-up of magnetic field in the process resembling sunspot activity.

\section{Waves in Shallow MHD}
\setcounter{equation}{0}
\label{Sect: Waves}

The MHD of stratified incompressible Boussinesq conducting fluid layer is described by the equations
\begin{eqnarray*}\label{MHD}
&{\bf V}_t+({\bf V}\cdot\nabla){\bf V}+{\bf f}\times{\bf V}=
      -\nabla {\Phi} + ({\bf B}\cdot\nabla){\bf B}+\varrho{\bf g}/\varrho_\ast,\\
&\varrho_t+{\bf V}\cdot\nabla\varrho=0,\quad
\nabla\cdot{\bf V}=\nabla\cdot{\bf B}=0,\\
&{\bf B}_t+({\bf V}\cdot\nabla){\bf B}=({\bf B}\cdot\nabla){\bf V},
\end{eqnarray*}
written in the local Cartesian coordinates of tangent plane layer ($x-$east, $y-$north, $z-$up), gravity and Coriolis forces included, ${\bf g}=[0,0,-g]$ and ${\bf f}=[0,0,f(y)]$. 
The fluid velocity ${\bf V}$, the magnetic field ${\bf B}$, the fluid density $\varrho$, and the divided by $\varrho_\ast$ pressure ${\Phi}$ are unknown functions of $x,y,z$, and time $t$; $\varrho_\ast$ is some value of almost constant density $\varrho$. Besides the thermodynamic pressure, ${\Phi}$ also includes the magnetic pressure and the term resulting from the centrifugal force. The magnetic field is normalized to have velocity units 
(${\bf B}\sqrt{\varrho_\ast\mu}$ is the real magnetic field, $\mu$ is magnetic permeability); the magnetic diffusion and the viscous force are neglected.

This dynamics has steady solution 
$${\bf V}_0=0,\; {\bf B}_0=[B_0, 0, 0],\; \varrho_0(z),\; \Phi_0(z) \qquad (\Phi_0'=-g\varrho_0/\varrho_\ast).$$
Linearization on its background, ${\bf V}={\bf V}_0+{\bf v}, {\bf B}={\bf B}_0+{\bf b},\varrho=\varrho_0+\rho,$ 
and ${\Phi}={\Phi}_0+{\phi},$ gives 
\begin{subequations}\label{lnzn}
\begin{eqnarray}
&&v^x_t - f v^y =-{\phi}_x+B_0 b^x_x,\\
&&v^y_t + f v^x=-{\phi}_y+B_0 b^y_x,\\
&&v^z_t=-{\phi}_z-g\rho/\varrho_\ast +B_0 b^z_x,\\
&&\rho_t+v^z\varrho_0'=0,\quad \nabla\cdot{\bf v}=\nabla\cdot{\bf b}=0,\\
&&b^x_t = B_0 \, v^x_x,\quad b^y_t = B_0 \, v^y_x,\quad b^z_t=B_0 \, v^z_x,
\end{eqnarray}
\end{subequations}
cf. \cite{Brag87,Brag98}; subscripts $x,y,z,t$ denote partial derivatives, while superscripts $x,y,z$ denote vector components.

Take $\partial_x$ of (\ref{lnzn}b) and subtract $\partial_y$ of (\ref{lnzn}a)
\begin{subequations}
\begin{eqnarray}\label{vorticityEq}
(v^y_x-v^x_y)_t+f\left(\rho_t/\varrho_0'\right)_z+\beta v^y
=B_0 (b^y_x-b^x_y)_x;\quad
\end{eqnarray}
here we also use (\ref{lnzn}d); $\beta=f'(y)$. 
Similar, from the first two equations in (\ref{lnzn}e),
\begin{eqnarray}\label{jzEq}
(b^y_x-b^x_y)_t = B_0 (v^y_x-v^x_y)_x.
\end{eqnarray}
\end{subequations}
Now let us use the quasigeostrophic and hydrostatic approximations, 
common in Geophysical fluid dynamics, e.g. \cite{Va}
\begin{eqnarray*}
\qquad v^x=-{\phi}_y/f,\quad v^y={\phi}_x/f,\quad {\phi}_z=-g\,\rho/\varrho_\ast\,;\quad
\end{eqnarray*}
then  the equations (\ref{vorticityEq}) and (\ref{jzEq}) become closed system on the pressure ${\phi}$ 
and the vertical component of the electric current density $m=b^y_x-b^x_y$
\begin{subequations}\label{sys}
\begin{eqnarray}
&\left[\Delta {\phi} - \frac{f^2}{g}\left(\frac{{\phi}_z}{\varrho_0'/\varrho_\ast}\right)_{\hspace{-0.15cm}z}\right]_t+\beta {\phi}_x=fB_0 {m}_x,\\
&{m}_t=f^{-1}B_0 \Delta \phi_x
\end{eqnarray}
($\Delta$ denotes the 2D Laplacian $\partial_x^2+\partial_y^2$). These equations should be supplemented by the boundary condition of vanishing normal velocity at the core-mantle boundary
\begin{eqnarray}
v^z=0 \qquad \Leftrightarrow \qquad \phi_{z t}=0.
\end{eqnarray}
\end{subequations}

The system (\ref{sys}) conserves the following positive-definite ($\varrho_0'<0$) integral
\begin{eqnarray}\label{Enstrophy}
\int \left[\left(\frac{\Delta\phi}{f}\right)^{\hspace{-0.15cm}2}-
\frac{(\phi_{xz})^2+(\phi_{yz})^2}{g\varrho_0'(z)/\varrho_\ast}+{m}^2\right]\hspace{-0.1cm}dxdydz\quad
\end{eqnarray}
(the integration being over the fluid domain $z<0$).

The system (\ref{sys}) allows separation of variables 
\begin{eqnarray*}
\phi=Z(z)\breve{\phi}(x,y,t),\; {m}=Z(z) \breve{m}(x,y,t):\\
\frac{g}{f^2}\frac{\Delta\breve\phi_t+\beta\breve\phi_x-f B_0\breve m_x}{\breve\phi_t}=
\frac{1}{Z}\left(\frac{Z'}{\varrho_0'/\varrho_\ast}\right)'.
\end{eqnarray*}
Denoting the separation constant by $1/h$, we find
\begin{subequations}\label{separation}
\begin{eqnarray}
&\Delta\breve\phi_t -(f^2/gh)\breve\phi_t+\beta\breve\phi_x=f B_0 \breve m_x,\\
&\
\breve m_t=f^{-1}B_0 \Delta \breve\psi_x,\\
&\left(\frac{Z'}{\varrho_0'/\varrho_\ast}\right)'=\frac{1}{h} Z,\quad Z'(0)=Z(-\infty)=0.
\end{eqnarray}
\end{subequations}
Equations (\ref{separation}ab) imply dispersion relation
\begin{eqnarray}\label{DispRel}
\omega^2\left(k^2+{\sf r}^{-2}\right)+\beta p\,\omega=B_0^2 p^2 k^2;
\end{eqnarray}
The Rossby radius ${\sf r}$ and the wave vector ${\bf k}=(p,q)$ are introduced in Section 1.

\paragraph{Model examples.}
Exponential density profile:
\begin{subequations}
\begin{eqnarray}\label{ExpPrfl}
\varrho_0=\varrho_\ast\left[1-\epsilon \exp(z/H)\right],
\end{eqnarray}
$H$ is the effective depth, and the parameter $\epsilon\ll 1$ controls the total density variation.
Linear density profile:
\begin{eqnarray}\label{LinPrfl}
\varrho_0=\varrho_\ast\left[1-\epsilon(H+z)/H\right]
\end{eqnarray}
($z=-\infty$ is replaced by  $z=-H$, $-H<z<0$). In this case, $Z''+(\epsilon/H h)Z=0,\; Z'(0)=Z(-H)=0$, and so,
$h=4\epsilon H/(\pi n)^2, \; Z=\cos (n\pi z/2 H)\quad (n=1,3,5,\ldots)$.
More generally,
\begin{eqnarray}\label{general}
\varrho_0=\varrho_\ast\left[1-\epsilon \chi(z/H)\right]
\end{eqnarray}
\end{subequations}
with some function $\chi(\zeta)$ that monotonically increases from $0$ at $\zeta=-\infty$ to 1 at $\zeta=0$. Then $h=O(\epsilon H)$.

\medskip

If the magnetic field is negligible, 
we have the usual hydrodynamic Rossby waves, $\omega_R\approx-\beta p/(k^2+{\sf r}^{-2})$.
When $h=\infty$, the relation (\ref{DispRel}) coincides with the one derived by \cite{Hide66} for the 2D dynamics of the Taylor-Proudman columns extending through the bulk of the liquid core and constrained by the core-mantle boundary.
The dispersion relation (\ref{DispRel}) was obtained  for waves in the ``Shallow water'' MHD \cite{gilman} by \cite{ZaqOliBalShe}; in that case, $h$ is the depth of the shallow layer, and $g$ is the reduced gravity.
For large $k$,  the relation (\ref{DispRel}) gives the Alfven waves, $\omega\approx\pm B_0 p$.
 
\paragraph{Let us estimate the two scales (\ref{rl}) for the Earth and Sun.}
\underline{Earth:} $f\approx 7\times10^{-5}$ s$^{-1}$, 
$\beta\approx4\times10^{-11}$ m$^{-1}$ s$^{-1}$ (at latitude $30^\circ$).
The toroidal magnetic field  has no signature on the surface of the earth, and so, $B_0$ is somewhat uncertain; we assume $B_0\sim0.3$ m/s, while $c_g\sim 7$ m/s.
Then ${\sf r}\sim{\ell}\sim 90$ km.\\
\underline{Sun:} For the solar tachocline, the situation is less clear. The assumptions made to derive the system (\ref{sys}) can be unsatisfactory: The Boussinesq approximation can fail. The geostrophic balance for the Sun
is ``marginal'' \cite{Gilman14}. Moreover, the parameters of the tachocline significantly vary across it. Even the parameters $f$ and $\beta$ are less obvious, because of Sun's differential rotation. 
We assume $f\sim3\times10^{-6}$ s$^{-1}$, $\beta\sim10^{-14}$ m$^{-1}$s$^{-1}$, $B_0\sim c_g\sim300$m/s, see \cite{Schecter01}. Then ${\sf r}\sim3\times10^8$m and ${\ell}\sim 2\times10^8$m.

\begin{figure}
\includegraphics[width=\columnwidth]{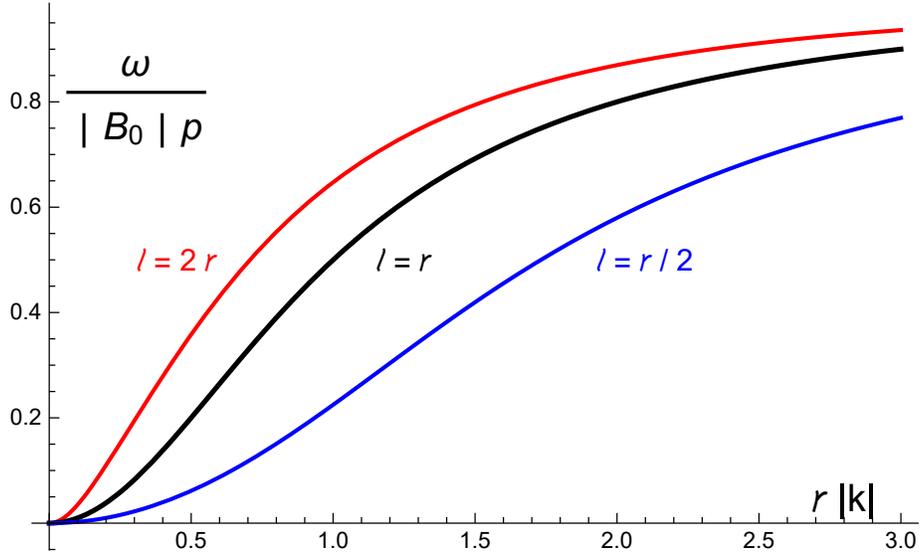}
\caption{The exact smaller (by absolute value) root of the equation (\ref{DispRel})
in three cases of relations between $\ell$ and ${\sf r}$.
When ${\ell}={\sf r}$, the exact root coincides with the Rossby form (\ref{magneticAlfven}b).}
\label{fig:RossbyApprox}
\end{figure}

Interestingly, in both cases (Earth and Sun), the scales $\sf r$ and ${\ell}$ are similar. 
If these scales are exactly equal 
\begin{eqnarray}\label{ScaleResonance}
c_g^2/f^2=|B_0/\beta|, 
\end{eqnarray}
the dispersion relation (\ref{DispRel}) can be factorized, and its two roots are given {\it exactly} by formulas (\ref{magneticAlfven}). They represent west-propagating Alfven wave and east-propagating magnetic Rossby wave 
(east is defined in the direction of rotation, north is $\pi/2$ counter-clock-wise turn from east, and so, $\beta>0$). 
The dispersion law (\ref{magneticAlfven}b) differs from the well known Rossby wave dispersion law only by a Doppler shift; but the parameters are different. 
Figure \ref{fig:RossbyApprox} shows the exact smaller frequency defined by the dispersion relation (\ref{DispRel}).

\section{Dynamo action}
\setcounter{equation}{0}
\label{Sect: Dynamo}
Consider energy share of each variable:
\begin{eqnarray*}
\left[\begin{array}{ccc}E_v^x&E_v^y&E_v^z\\ &E_\rho&\\ E_b^x&E_b^y&E_b^z\end{array}\right]\equiv
\frac{1}{2}\int_{-\infty}^0 \left[\begin{array}{ccc}|v^x|^2 &|v^y|^2 &|v^z|^2\\ &-g|\rho|^2/\varrho_\ast \varrho_0'&\\
|b^x|^2&|b^y|^2&|b^z|^2\end{array}\right] d z.
\end{eqnarray*}
One can check that the integral 
\begin{eqnarray*}\label{Energy}
E=\int (E_v^x+E_v^y+E_v^z+E_\rho+E_b^x+E_b^y+E_b^z)d x d y
\end{eqnarray*}
is conserved by the system (\ref{lnzn}) under the boundary condition (\ref{sys}c).
This conservation holds because the vector ${\bf B}_0$ has no $z$-component; otherwise the energy integral would need to be extended outside of the fluid domain.

We can express all variables via $\phi$ and ${m}$: The quasigestrophy gives $v^x$ and $v^y$;  the hydrostacy gives $\rho$. By the first equation in (\ref{lnzn}d), $v^z=-\rho_t/\varrho_0$.
From the first equation in (\ref{lnzn}d) and the third equation in (\ref{lnzn}e), 
\begin{eqnarray*}
\left(b^z+B_0\rho_x/\varrho_0' \right)_t=0
\;\Rightarrow\;
b^z=-B_0\rho_x/\varrho_0'.
\end{eqnarray*}
This is the linearized version of the magnetic analogue of Ertel's theorem \cite{Hide83}, 
saying that the quantity ${\bf B}\cdot\nabla\varrho$ --- in the fully nonlinear dynamics --- 
is constant along the lines of fluid motion.
Using the last equation in (\ref{lnzn}d) and the definition of ${m}$, we find $b^x$ and $b^y$
\begin{eqnarray*}
\Delta b^x=-{m}_y-b^z_{z x},\quad \Delta b^y={m}_x-b^z_{z y}.
\end{eqnarray*}
Take a single wave 
$$\phi=Z(z) e^{i(p x + q y - \omega t)}.$$
According to the equations (\ref{sys}a) and (\ref{separation}c),
$${m}=\frac{\omega(k^2+{\sf r}^{-2})+\beta p}{f\, B_0\; p} Z e^{i(p x + q y - \omega t)}.$$
Now we find all other variables in that wave 
[the right-hand-sides in the following equations are to be multiplied by $\exp{\{i(p x + q y - \omega t)\}}$ ]
\begin{subequations}\label{polarization}
\begin{eqnarray}
&&v^x=-\frac{i q}{f} Z,\quad v^y=\frac{i p}{f}Z, \quad \rho=-\frac{\varrho\ast}{g} Z',\\
&&v^z=\frac{-i \omega Z'}{g\varrho_0'/\varrho_\ast},\qquad
b^z=\frac{i p B_0 Z'}{g\varrho_0'/\varrho_\ast},\\
&&b^x=\frac{1}{k^2}\left[i q \frac{\omega(k^2+{\sf r}^{-2})+\beta p}{f B_0 p}\,-\,\frac{p^2 B_0}{g h}\right]Z,\\
&&b^x=\frac{1}{k^2}\left[-i p \frac{\omega(k^2+{\sf r}^{-2})+\beta p}{f B_0 p}\,-\,\frac{p q B_0}{g h}\right]Z,\quad
\end{eqnarray}
\end{subequations}
[expressions (\ref{polarization}cd) use (\ref{separation}c)].

The formulas (\ref{polarization}) determine the energy shares; we write them for the magnetic Rossby wave, when $\omega$ is given by (\ref{magneticAlfven}b), and the relation (\ref{ScaleResonance}) holds
\begin{subequations}\label{Share}
\begin{eqnarray}
&&E_v^x=\frac{q^2}{f^2},\quad E_v^y=\frac{p^2}{f^2},\quad E_\rho=\frac{1}{c_g^2},\\
&&E_v^z=\frac{\beta^2 p^2 k^4}{f^4 (k^2+{\sf r}^{-2})^2} \Lambda^2,\quad 
    E_b^z=\frac{\beta^2 p^2}{f^4} \Lambda^2,\\
&&E_b^x=\frac{1}{f^2 k^4}\left[q^2(k^2+{\sf r}^{-2})^2+p^4(\beta/f)^2\right],\\
&&E_b^y=\frac{1}{f^2 k^4}\left[p^2(k^2+{\sf r}^{-2})^2+p^2 q^2(\beta/f)^2\right],
\end{eqnarray}
\end{subequations}
where all right-hand-sides should be multiplied by $(1/2)\int_{-\infty}^0 Z^2 dz$; and
\begin{eqnarray*}\label{zeta}
\Lambda^2=h^2 \left.\int \left(\frac{Z'}{\varrho_0'/\varrho_\ast}\right)^{\hspace{-0.1cm}2} dz\right/\int Z^2 dz.
\end{eqnarray*}

The formulas (\ref{Share}) show that the content of magnetic energy $E_b=E_b^x+E_b^y+E_b^z$ is bigger in longer waves. (More on this later.) We assume that sources generate relatively short waves. Interacting between each other, they produce longer waves. Most of the energy is carried by long waves. The energy accumulation in long waves is a very general fact, occurring in a majority of physical systems. It is independent of cascade: Even if the energy cascade exists, it could be well directed towards the smaller scales. 
For example, in the weakly nonlinear system of gravity waves \cite{ZakhKS}, the energy does cascade towards the small scales, but the long waves carry most of the energy. Specifically in the system of slow magnetic waves, the energy accumulation in long waves was indicated \cite{apjBalk14} by the infrared divergence of the corresponding Kolmogorov-Zakharov spectrum.

However, this transfer of the small-scale kinetic energy into the large-scale magnetic energy is insufficient for dynamo.  The above approach requires dominating zonal magnetic field ${\bf B_0}$; but it will gradually decay (since ${\bf B}_0$ slowly varies in space). There should be some mechanism of energy supply specifically into large-scale zonal magnetic field. This mechanism --- described below --- is due to the extra invariant.

The main interactions in an arbitrary wave system are due to the resonances
\begin{eqnarray}\label{Resonance}
p_1=p_2+p_3,\quad q_1=q_2+q_3,\quad\omega_1=\omega_2+\omega_3,
\end{eqnarray}
where $\omega_i=\omega({\bf k}_i)$ is the frequency of the wave with wave vector ${\bf k}_i=(p_i,q_i)\; (i=1,2,3)$.

A dispersion law $\omega({\bf k})$ is said to be {\it degenerative}
\cite{ZSch0} if there exists an independent function $\eta({\bf k})$ conserved in the resonance interactions: 
Whenever vectors ${\bf k}_1, {\bf k}_2, {\bf k}_3$ satisfy the relations (\ref{Resonance}), 
we also have  $\eta({\bf k}_1)=\eta({\bf k}_2)+\eta({\bf k}_3)$;
the function  $\eta({\bf k})$ should be linearly independent of the functions
$p,\,q$ and $\omega({\bf k})$ --- otherwise the latter equation 
is a mere linear combination of the resonance equations (\ref{Resonance}).

In a weakly nonlinear system, the conservation of the function $\eta({\bf k})$ in resonance interactions (\ref{Resonance}) implies the adiabatic conservation of the integral
\begin{eqnarray}\label{Invariant}
I=\int \frac{\eta({\bf k})}{\omega({\bf k})}{\mathcal E}_{\bf k} d{\bf k} \quad 
[{\mathcal E}_{\bf k} \mbox{ is the energy spectrum}].\quad
\end{eqnarray}
 The existence of the extra invariant is independent of the form of nonlinearity, as long as the system is Hamiltonian (which is usually the case for a physical system). This is proved by showing cancellation of the small denominator \cite{ZSch}.

The conservation is {\it adiabatic} in the sense that it holds approximately over long time: If wave amplitudes are of the order $\epsilon\rightarrow0$, then $I=O(\epsilon^2)$, but $\Delta I=I(t)-I(0)$ is at least $O(\epsilon^3)$ over time intervals  $O(\epsilon^{-1})$.

The Rossby wave dispersion law (\ref{magneticAlfven}b) is degenerative \cite{BNZ,B1991}, with the function (\ref{ETA}). This is actually the only independent function conserved in the interactions (\ref{Resonance}). 
The conservation of the corresponding integral (\ref{Invariant}) implies the energy accumulation in the region $p\ll q$ \cite{B2005}. We will explain this fact and show the energy accumulation in the large-scale zonal magnetic field. 
Consider a modification of $\eta$ 
\begin{eqnarray}\label{mETA}
\tilde\eta({\bf k})=|B_0|\,\eta({\bf k})\,+\,
                             2\sqrt{3}\;{\sf r}\,[\,\omega({\bf k})-|B_0|\, p\,]. 
\end{eqnarray}
Obviously, $\tilde\eta$ is also conserved in the resonance interactions (it is a linear combination of conserved functions), but it has better asymptotic properties
\begin{eqnarray}\label{Asymptotics}
\tilde\eta({\bf k})\sim|B_0|\,8\sqrt{3}\times\left\{\begin{array}{ll}
p^3 \;\frac{{\sf r}}{q^2(1+{\sf r}^2q^2)^3}, &\; p\rightarrow 0,\\
k^{-5} \; \frac{p^3(p^2+5 q^2)}{5{\sf r}^5 k^5}, &\; k\rightarrow\infty,
\end{array}\right.
\end{eqnarray}
so that, $\tilde\eta$ rapidly decreases as $p\rightarrow0$ or $k\rightarrow\infty$ (faster than $\omega$, $p$, or $\eta$ alone). In (\ref{Asymptotics}), the first asymptotics holds for any $q$, and the second --- in all directions.

A wave system conserves the momentum and energy
\begin{subequations}
\begin{eqnarray*}\label{EnergyMomentum}
P^x=\int \frac{p}{\omega}\,{\mathcal E} d{\bf k}, \quad
P^y=\int \frac{q}{\omega}\,{\mathcal E} d{\bf k}, \quad
E=\int {\mathcal E} d{\bf k}.
\end{eqnarray*}
The conservation of these invariants are implied by the resonance relations (\ref{Resonance}).
For the Rossby system, the $y$-momentum $P^y$ is not a real physical invariant (see \cite{BaYo}), since the function $q/\omega$ is singular (as $p\rightarrow0$).
The linear combination of the $x$-momentum and the energy determines the enstrophy
\begin{eqnarray*}
F={\sf r}^2\,(\,|B_0|\,P^x\,-\,E\,)=\int \frac{{\mathcal E}_{\bf k}}{k^2}\,d{\bf k}.
\end{eqnarray*}
\end{subequations}
The conservation of the energy $E$ and enstrophy $F$ implies that most of the energy from the source should be transferred towards large $k$, and most of the enstrophy --- towards small $k$.

\begin{figure}
\includegraphics[width=\columnwidth]{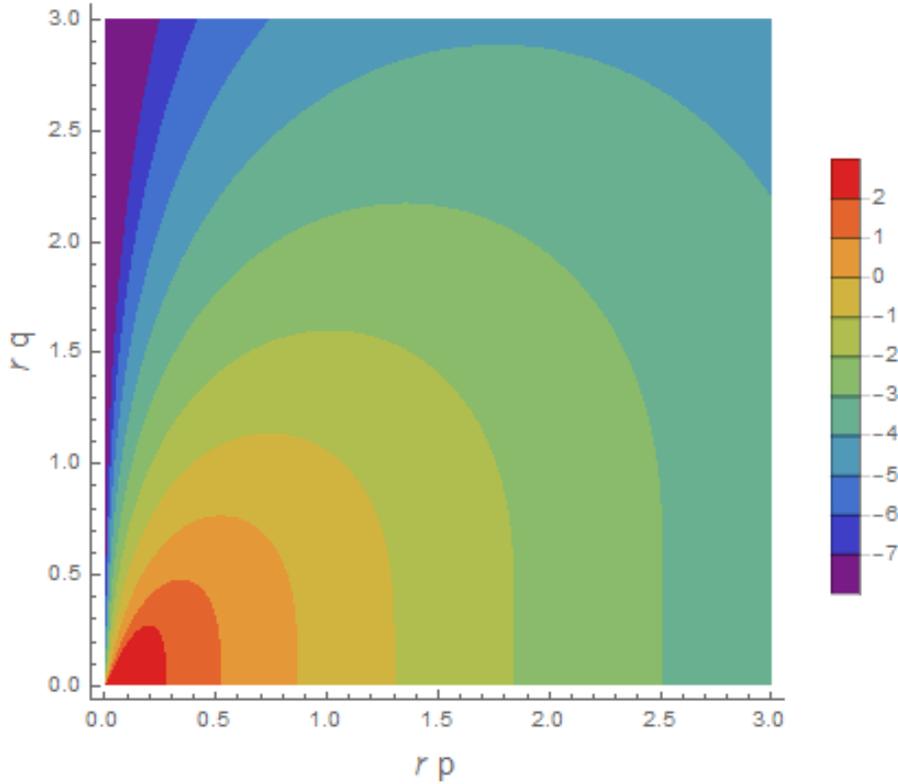}
\caption{Extra invariant per enstrophy, i.e.\ the function (\ref{ratio}). Color represents the values of $\ln[G(p,q)]$.
Each of the two extreme areas (where $G<e^{-7}$ or $G>e^2$) is shown by a single color.}
\label{fig:ContourPlot}
\end{figure}

The energy parcel ${\mathcal E}_{\bf k}d{\bf k}$ always carries with it the enstrophy parcel
$k^{-2}{\mathcal E}_{\bf k}d{\bf k}$ and the parcel
$[\tilde\eta({\bf k})/\omega({\bf k})]{\mathcal E}_{\bf k}d{\bf k}$ of the extra invariant 
$\tilde I=\int (\tilde\eta/\omega){\mathcal E} d{\bf k}$, the latter being similar to the integral (\ref{Invariant}).
Figure \ref{fig:ContourPlot} shows the ratio of the extra invariant parcel to the enstrophy parcel
\begin{eqnarray}\label{ratio}
G(p,q)=\left.\frac{\tilde\eta({\bf k})}{\omega({\bf k})}{\mathcal E}_{\bf k}d{\bf k}\right/
\frac{1}{k^2}{\mathcal E}_{\bf k}d{\bf k}=\frac{k^2\tilde\eta({\bf k})}{\omega({\bf k})}.
\end{eqnarray}
The Fig.\ \ref{fig:ContourPlot} and asymptotics (\ref{Asymptotics}) imply the energy accumulation in region $p\ll q$. Indeed, the energy accumulation in large scales $\mathcal L$ (${\sf r}k\ll 1$) would mean that a significant amount of enstrophy is brought to  this region. 
If the energy is accumulated at large scales away from the region $p\ll q$, then a huge amount of the extra invariant would be required. This is because the Fig.\ \ref{fig:ContourPlot} shows large values of the function (\ref{ratio}) away from the $q$-axis, when ${\sf r}k\ll 1$. But the extra invariant generation is limited by the source.
Now, according to the equations (\ref{Share}), the accumulated energy is mostly $E_b^x$.

To see this, let us first note that the second term on the right in each of the equations (\ref{Share}cd) is negligible. Indeed, these equations can be written in the form
\begin{eqnarray*}
E_b^x=\frac{q^2}{f^2}\left(\frac{k^2+{\sf r}^{-2}}{k^2}\right)^{\hspace{-0.1cm}2}
\left[1+\left(\frac{p^2}{k^2+{\sf r}^{-2}}\frac{\beta{\mathcal L}^y}{f}\right)^{\hspace{-0.1cm}2}\right],\\
E_b^y=\frac{p^2}{f^2}\left(\frac{k^2+{\sf r}^{-2}}{k^2}\right)^{\hspace{-0.1cm}2}
\left[1+\left(\frac{q^2}{k^2+{\sf r}^{-2}}\frac{\beta{\mathcal L}^y}{f}\right)^{\hspace{-0.1cm}2}\right],
\end{eqnarray*}
where ${\mathcal L}^y=q^{-1}$ is the length scale in the $y$-drection. 
The applicability of the $\beta$-plane approximation requires $\beta {\mathcal L}^y\ll f$, 
and so, the brackets here are roughly 1.  Thus,
\begin{eqnarray*}
\frac{E_v^x}{E_b^x}\sim\frac{E_v^y}{E_b^y}\sim\frac{E_v^z}{E_b^z}=
\left(\frac{{\sf r}^2 k^2}{1+{\sf r}^2 k^2}\right)^{\hspace{-0.1cm}2} \sim\left\{\begin{array}{lr} 1, & {\sf r} k\gtrsim1,\\({\sf r} k)^4, & \;{\sf r} k\ll 1.\end{array}\right.
\end{eqnarray*}
(Strictly speaking, when ${\sf r}k\gtrsim1$, the value of expression ${\sf r}^2 k^2/(1+{\sf r}^2 k^2)$ changes from 1/2 to 1, but we disregard this.) By the third equation in (\ref{Share}a) and relation (\ref{ScaleResonance}),
\begin{eqnarray*}
\frac{E_\rho}{E_b^x+E_b^y}\sim\left(\frac{{\sf r} k}{1+{\sf r}^2 k^2}\right)^{\hspace{-0.1cm}2} \sim\left\{\begin{array}{lr} ({\sf r} k)^{-2},& {\sf r}k\gg1,\\ 1/4,& {\sf r} k\sim1,\\({\sf r} k)^2, & \;{\sf r} k\ll 1.\end{array}\right.
\end{eqnarray*}
These asymptotics show that in short waves (${\sf r}k\gtrsim1$), the magnetic and mechanical energies are of the same order, but long waves (${\sf r}k\ll1$) carry mostly the magnetic energy. Now,
\begin{subequations}\label{domination}
\begin{eqnarray}
\frac{E_b^y}{E_b^x}&\sim& \left(\frac{p}{q}\right)^{\hspace{-0.1cm}2},\\
\frac{E_b^z}{E_b^x}&\sim& \left(\frac{p}{q}\right)^{\hspace{-0.1cm}2}
\left(\frac{{\sf r}^2 k^2}{1+{\sf r}^2 k^2}\right)^{\hspace{-0.1cm}2}
\left(\frac{\beta \Lambda}{f}\right)^{\hspace{-0.1cm}2}.
\end{eqnarray}
\end{subequations}
For the density profile (\ref{general}), the distance $\Lambda$ has magnitude of the effective depth $H\lesssim L$, 
and so $\beta \Lambda\lesssim f$. Therefore, the expressions (\ref{domination}) show domination of $E_b^x$ in the large-scale magnetic energy.

Let us note that the relation (\ref{ScaleResonance}) is dependent on latitude $\vartheta$, since
\begin{eqnarray*}
f=2\Omega_0\sin\vartheta,\qquad \beta=(2\Omega_0/R_0)\cos\vartheta
\end{eqnarray*}
($\Omega_0$ is the angular speed of Earth's rotation); $f\rightarrow0$ as $\vartheta\rightarrow0$, and  $\beta\rightarrow0$ as $\vartheta\rightarrow\pi/2$.
So, the relation (\ref{ScaleResonance}) --- which we can refer to as {\it the scale resonance} ---  is realized at some latitude $\vartheta$ (provided $c_g$ does not strongly depend on latitude, as $c_g$ is not directly related to rotation or magnetic field). Because of the energy accumulation in the large-scale zonal magnetic field, the magnitude $B_0$ will increase, and therefore the relation (\ref{ScaleResonance}) will be no longer satisfied at the latitude $\vartheta$. 
But  the scale resonance (\ref{ScaleResonance}) will be satisfied 
at a new {\it lower} latitude $\vartheta-d\vartheta$, and so, the magnetic field will be still maintained. If the energy supply is large enough, $B_0$ will still increase. 
The latitude responsible for the generation of magnetic field will gradually decrease. 
This resembles sunspot activity; a similar phenomenon should take place in the Earth.
In this process, the basic zonal magnetic field is build-up to 
\begin{eqnarray}\label{build-up}
|B_0|\approx\frac{c_g^2}{2 R_0 \Omega_0} \frac{\cos\vartheta}{\sin^2\vartheta}.
\end{eqnarray}
As the resonance latitude [where the scale resonance (\ref{ScaleResonance}) is realized] changes, say from $35^\circ$ to $15^\circ$, the Alfven speed $B_0$ increases about 6 times (for Earth, from 0.25m/s to 1.5m/s; for Sun, from 75m/s to 450m/s), provided $c_g$ remains constant.

\bibliographystyle{unsrt}
\bibliography{My}
\end{document}